\documentclass{article}
\usepackage{template/spconf,amsmath,amsfonts,epsfig}
\usepackage{hyperref}
\usepackage{multirow}
\usepackage{graphicx}
\usepackage{float}
\usepackage{fancyhdr}
\usepackage{cite}
\usepackage{booktabs}
\let\OLDthebibliography\thebibliography
\renewcommand\thebibliography[1]{
  \OLDthebibliography{#1}
  \setlength{\parskip}{0pt}
  \setlength{\itemsep}{0pt plus 0.3ex}
}

\fancypagestyle{FirstPage}{
}

\begin{document}\sloppy

\begin{titlepage}
  \centering
  This document is the accepted version of the paper that has been published as:
N. Le et al., "Bridging the Gap Between Image Coding for Machines and Humans," 2022 IEEE International Conference on Image Processing (ICIP), Bordeaux, France, 2022, pp. 3411-3415, doi: 10.1109/ICIP46576.2022.9897916, url: https://ieeexplore.ieee.org/document/9897916.
© 2022 IEEE. Personal use of this material is permitted. Permission from IEEE must be obtained for all other uses, in any current or future media, including reprinting/republishing this material for advertising or promotional purposes, creating new collective works, for resale or redistribution to servers or lists, or reuse of any copyrighted component of this work in other works.
\end{titlepage}

\title{
Bridging the gap between image coding for machines and humans
}

\name{\begin{tabular}{c}
Nam Le$^{\ast\dag}$ \qquad
Honglei Zhang$^{\dag}$ \qquad
Francesco Cricri$^{\dag}$ \qquad
Ramin G. Youvalari$^{\dag}$ \\
Hamed Rezazadegan Tavakoli$^{\dag}$ \qquad 
Emre Aksu$^{\dag}$ \qquad 
Miska M. Hannuksela$^{\dag}$ \qquad
Esa Rahtu$^{\ast}$
\end{tabular}}
\address{$^{\dag}$Nokia Technologies, $^{\ast}$Tampere University 
}

\maketitle
\newcommand{\fullresincluded}{}

\newcommand{\newparagraph}[1]{\par\textbf{#1}}

\newcommand{\Tensor}[1]{\boldsymbol{#1}}
\newcommand{\Loss}[1]{\mathcal{L}_{#1}}
\newcommand{\LossFT}[1]{\bar{\mathcal{L}}_{#1}} 
\newcommand{\Wof}[1]{\bar{w}_{#1}} 
\newcommand{\Model}[2]{\boldsymbol{#1}({#2})} 
\newcommand{\ModelWeights}[1]{\boldsymbol{\theta}_{\boldsymbol{#1}}} 
\newcommand{\Wmodel}[2]{\Model{#1}{{#2};\ModelWeights{#1}}} 
\newcommand{\goodresult}[1]{\textcolor{green}{#1}}
\newcommand{\badresult}[1]{\textcolor{red}{#1}}

\renewcommand*{\figureautorefname}{Fig.}
\newcommand{\etal}{\textit{et al.}}

\newcommand{\litencoder}{E}
\newcommand{\litdecoder}{D}
\newcommand{\litprobmodel}{P}
\newcommand{\litquantizer}{Q}
\newcommand{\litimg}{x}
\newcommand{\litlatent}{y}
\newcommand{\lithyperlatent}{z}
\newcommand{\litprior}{p}
\newcommand{\litweight}{w}
\newcommand{\litweightset}{W}
\newcommand{\litlossrate}{rate}
\newcommand{\litlosstask}{task}
\newcommand{\litlossmse}{mse}
\newcommand{\litlossfinetune}{total}
\newcommand{\litlossperceptual}{proxy}
\newcommand{\img}{\Tensor{\litimg}}
\newcommand{\resimg}{\Tensor{\hat{\litimg}}} 
\newcommand{\latent}{\Tensor{\litlatent}}
\newcommand{\iqlatent}{\Tensor{\hat{\litlatent}}} 
\newcommand{\tqlatent}{\Tensor{\Tilde{\litlatent}}} 
\newcommand{\hyperlatent}{\Tensor{\lithyperlatent}}
\newcommand{\tqhyperlatent}{\Tensor{\Tilde{\lithyperlatent}}}
\newcommand{\iqhyperlatent}{\Tensor{\hat{\lithyperlatent}}}
\newcommand{\prior}[1]{\litprior_{#1}}
\newcommand{\encoder}[1]{\Wmodel{\litencoder}{#1}}
\newcommand{\decoder}[1]{\Wmodel{\litdecoder}{#1}}
\newcommand{\probmodel}[1]{\Wmodel{\litquantizer}{#1}}
\newcommand{\wrate}{\litweight_{\litlossrate}}
\newcommand{\wtask}{\litweight_{\litlosstask}}
\newcommand{\wmse}{\litweight_{\litlossmse}}
\newcommand{\wset}{\mathcal{\litweightset}}
\newcommand{\lossrate}{\Loss{\litlossrate}}
\newcommand{\losstask}{\Loss{\litlosstask}}
\newcommand{\lossmse}{\Loss{\litlossmse}}
\newcommand{\lossperceptual}{\Loss{\litlossperceptual}}
\newcommand{\lossproxy}{\lossperceptual}

\newcommand{\gen}{\mathcal{G}}
\newcommand{\disc}{\mathcal{D}}
\newcommand{\lossLtwo}{\Loss{L2}}
\newcommand{\lossgan}{\Loss{GAN}}
\newcommand{\wadv}{\lambda_{a}}


\begin{abstract}


Image coding for machines (ICM) aims at reducing the bitrate required to represent an image while minimizing the drop in machine vision analysis accuracy. In many use cases, such as surveillance, it is also important that the visual quality is not drastically deteriorated by the compression process. Recent works on using neural network (NN) based ICM codecs have shown significant coding gains against traditional methods; however, the decompressed images, especially at low bitrates, often contain checkerboard artifacts. We propose an effective decoder finetuning scheme based on adversarial training to significantly enhance the visual quality of ICM codecs, while preserving the machine analysis accuracy, without adding extra bitcost or parameters at the inference phase. The results show complete removal of the checkerboard artifacts at the negligible cost of $-1.6\%$ relative change in task performance score. In the cases where some amount of artifacts is tolerable, such as when machine consumption is the primary target, this technique can enhance both pixel-fidelity and feature-fidelity scores without losing task performance.

\end{abstract}
\begin{keywords}
    Image coding for machines, GANs, finetuning, VCM
\end{keywords}
\thispagestyle{FirstPage}
\section{Introduction}
\label{sec:intro}

Traditional image coding systems such as JPEG \cite{jpeg} and VVC \cite{cit:vvc} compress images to reduce the amount
of the data required for storing and transferring while maintaining
satisfactory visual quality for human viewers. On the other hand,
Image Coding for Machines (ICM) systems aim to provide better compression
efficiency when the primary consumers are machines that perform certain
computer vision tasks \cite{yang2021videocoding, VCM_duan2020videocoding}. Recent works on ICM have shown convincing evidences that machine-oriented image codecs significantly improve the compression
efficiency over traditional image codecs \cite{icassp_paper,proxy_loss_fischer2021,yang2021videocoding}.
In many applications, e.g. surveillance, although machines are 
the main consumers, human involvement is occasionally required or even mandatory. Thus, support for human consumption is desired for 
ICM codecs.

However, the outputs of convolutional neural network (CNN) based ICM codecs, in particular at low bit-rate range, usually contain repetitive artifact patterns,
also referred to as ``checkerboard artifacts'' \cite{proxy_loss_fischer2021,icassp_paper},
which could be easily perceived by humans as ``distortion'' and significantly
degrade the experience for human viewers. To support human consumption,
ICM systems often include separate branches for machine consumption
and human consumption \cite{feature_residuals_seppala2021,yang2021videocoding,VCM_duan2020videocoding}, which unavoidably incurs bitrate overheads due to the need for encoding two bitstreams. In addition, such architecture design increases the complexity of the codec significantly.

In this paper, we propose a novel training technique for end-to-end
learned ICM codecs to improve the visual quality without additional
processing components or bitstreams for human consumption. 
In the proposed technique, selected layers from the decoder of a
pretrained ICM codec are finetuned using a PatchGAN \cite{isola2017imagetoimage} adversarial training mechanism. Our experiments show that after being finetuned, the quality of the decoded
images is significantly improved without compromising
the performance of the machine tasks.


\section{Related work}
\label{sec:related-work}

End-to-end learned codecs are a promising direction over conventional codecs for machine consumption since the codec may be optimized directly via a learning signal obtained through the task networks performing computer vision tasks. The authors of \cite{icassp_paper, icme_paper} proposed an ICM codec with encoder, decoder, and probability model implemented by neural networks. An input image is first converted to a latent representation. Then, an arithmetic encoder encodes the quantized latent representation using the prior distribution estimated by the probability model. At the decoder side, an arithmetic decoder decodes the bitstream into the latent representation with the prior from the same probability model as the one on the encoder side and the ICM decoder reconstructs the image from the latent representation. 

Although machine consumption is the main target of an ICM codec, human consumption is required for many applications, e.g., surveillance systems. Various system architectures have been proposed to support this requirement. Seppala \etal \cite{feature_residuals_seppala2021} proposed a multi-layer coding system comprising two branches: one implemented by a conventional codec for human consumption and one  implemented by a learned codec that transfers additional information to enhance the performance of machine consumption. A similar architecture is proposed in \cite{choi2022scalable} where multiple learned codecs are used for human vision and various machine tasks separately. These designs  significantly complicate the system and make optimization difficult. Also, an additional bitstream is required for targeting human consumption of images. As the task networks of these ICM systems are trained using images which are either uncompressed or compressed with high fidelity for human vision, the outputs of the ICM codecs are visually acceptable for humans. However, clearly noticeable checkerboard artifacts are present in the output, especially at lower bitrates, which significantly degrade the quality perceived by human viewers. Thus, it is important to improve the quality of the outputs of ICM codecs for human consumption without significantly degrading the machine task. 
Zhang \etal \cite{blurpool_zhang2019making} propose to blur the features in CNN to mitigate the aliasing problem, but we found it is ineffective in deeper CNNs.

Generative Adversarial Networks (GANs) were introduced as a system for generating realistic natural images and later used in many other applications, including for visual quality enhancement. Isola \etal \cite{isola2017imagetoimage} introduced the PatchGAN technique for image-to-image translation, which restricts the attention of the networks in local image patches, thus enforces high-frequency correctness. Further, in \cite{kupyn2019deblurganv2}, both the global and local patches are used for image deblurring. GANs have been also used in video coding systems
. In \cite{mentzer2020highfidelity}, the authors used conditional GANs architecture to improve the perceptual visual quality of learned image compression targeting human consumption.
The analysis in \cite{fischer2021analysis} suggests that the said adversarial system improves
task performance of the codec.

Next, we will describe the proposed PatchGAN-based method that significantly improves the visual quality of the compressed images of an ICM codec for human consumption.

\section{Proposed method}
\label{sec:method}
\subsection{Baseline image codec model}
\label{ssec:basemodel}
Similar to \cite{icassp_paper, icme_paper,proxy_loss_fischer2021}, we first train a base codec using metrics such as mean squared error (MSE) for human consumption, then finetune it to achieve better task performance. 
We follow the overall model training strategy as described in \cite{icassp_paper}.


Our codec comprises 3 types of building blocks: \textbf{B}asic, \textbf{D}ownsampling, and \textbf{U}psampling. The \textbf{B} blocks are type A basic blocks defined in ResNet \cite{resnet} with 128 output channels and PReLU activation function \cite{prelu}. The \textbf{D} blocks are \textbf{B} blocks that have stride-2 \textit{conv} (\textit{convS2}) for their first \textit{conv} and shortcut projection. Similarly, the \textbf{U} blocks use a sequence of a 512-channel \textit{conv} followed by a PixelShuffle \cite{pixelshuffle} layer (\textit{PSConv}) for those parts. The layer orders in the encoder and decoder of the codec can be described as \textbf{D}-\textbf{B}-\textbf{D}-\textbf{B}-\textit{convS2} and \textbf{B}-\textbf{U}-\textbf{B}-\textbf{U}-\textbf{B}-\textit{PSConv}, respectively.


The training loss, which imposes high task performance, follows the same formulation in \cite{icassp_paper}:
\begin{equation}
   \Loss{total} = \wrate\lossrate + \wmse\lossmse + \wtask\losstask ,
\end{equation}
where $\wrate$, $\wmse$, $\wtask$ are scalar weights, $\lossrate$, $\lossmse$, $\losstask$ are the loss terms for bitcost, distortion and task performance, respectively. These terms are defined in the same way as in \cite{icassp_paper}, except for $\losstask$ which is replaced by a proxy loss $\lossproxy$ in our setup. Due to the correlations in the intermediate level features of different vision tasks \cite{icme_paper}, we can use an intermediate feature distortion metric as a proxy for $\losstask$, thus making the codec task-agnostic. Additionally, using a feature-based loss as such enables the training of the model with cropped images which is much more efficient. Similar to \cite{icme_paper,pfilter_ahonen2021}, we define $\lossproxy$ as follows:
\begin{equation}
   \lossproxy = \mathbf{MSE}(\mathcal{F}_2(\img), \mathcal{F}_2(\resimg))
   + \mathbf{MSE}(\mathcal{F}_4(\img), \mathcal{F}_4(\resimg)),
   \label{eq:proxy-loss}
\end{equation}
where $\mathbf{MSE}$ denotes the mean square error and $\mathcal{F}_k(\Tensor{t})$ denotes a feature tensor extracted from the $P_k$ layer of the FPN \cite{resnetfpn} feature extractor network given the input $\Tensor{t}$.

\subsection{Codec finetuning with PatchGAN}
\label{ssec:gansfinetuning}
\begin{figure}[ht]
   \begin{minipage}[b]{\linewidth}
      \centering
      \centerline{\includegraphics[width=\linewidth]{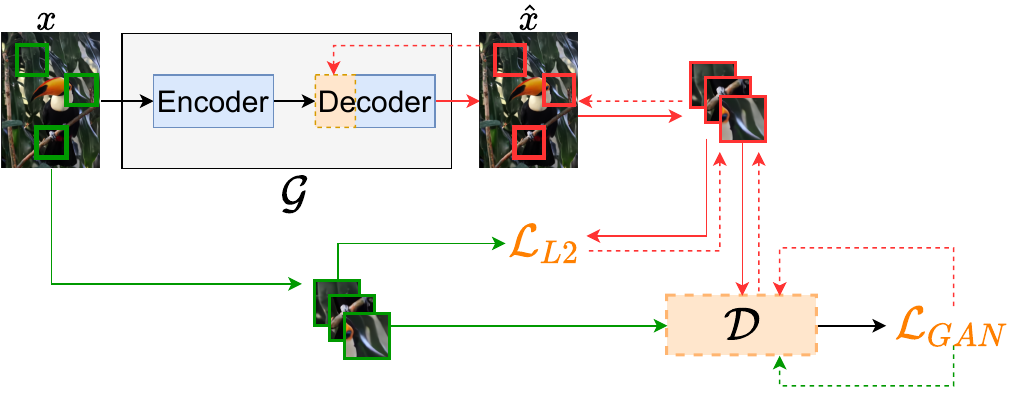}}
      
   \end{minipage}
   \caption{The overview of the finetuning scheme using PatchGAN discriminator. The dashed lines denote gradient back-propagation flows and dotted boxes denote the parameters getting updated by the optimizer. The green lines indicate the data from the input of the ICM codec and the red lines indicate the data from the output of the ICM codec.}
   \label{fig:overview}
\end{figure}

In order to improve the visual quality of a learned image codec, one can resort to re-training the model for human vision distortion metrics such as PSNR or SSIM \cite{ssim_wang2004image}. However as demonstrated in recent works  \cite{proxy_loss_fischer2021,fischer2021analysis}, high scores in these metrics usually come with the cost of significant degradation of task performance. Furthermore, \cite{icme_paper} provides us with evidences that even finetuning a single component of the system can achieve good performance gains. We propose a model finetuning scheme where a PatchGAN \cite{isola2017imagetoimage} discriminator is applied to guide the codec to effectively suppress the artifacts in its outputs. In this scheme, illustrated by \autoref{fig:overview}, the image codec acts as the generator $\gen$ that takes uncompressed image $\img$ as input and generates the compressed image $\resimg=\gen(\img)$, and the discriminator $\disc$, implemented as a CNN following \cite{isola2017imagetoimage}, tries to distinguish the image patches from $\img$ from the ones from $\resimg$ by detecting the noise pattern in $\resimg$ produced by the codec. The reasons for choosing  PatchGAN over normal GANs are: \textit{i)} a small patch from $\resimg$ can contain an adequate amount of artifacts; \textit{ii)} the training runs much faster with small patches; \textit{iii)} we can limit the random hallucinations that are typically introduced in GAN-based training, which may damage the task performance. Our experiments show that the compression artifacts are sufficiently removed by finetuning only the decoder. With the encoder frozen, the bitrate is unchanged.   

\newparagraph{Finetuning objective:} 
We empirically found the proposed system is sufficiently robust to different adversarial loss functions.
Therefore, we choose the vanilla GAN objective function for the sake of simplicity:
\begin{equation}
   \lossgan(\gen, \disc) = \mathbb{E}_{\img \sim p_{\img}} \left[\log \disc(\img)\right] +  \left[\log(1- \disc(\gen(\img)))\right]
\end{equation}
Unlike \cite{isola2017imagetoimage}, we only sample a small fixed number of square patches to feed the discriminator since the artifacts appear uniformly in the decoded images. Besides, we do not condition the discriminator on the input image $\img$. In our system, the generator is a pretrained codec that outputs a fair reconstruction $\resimg$ of $\img$. Therefore, we use $\ell_2$-distance to impose a weak pixel-fidelity constraint. Our funetuning objective is formulated as:
\begin{equation}
      G^{*} = \mathrm{arg}~\underset{\gen}{\mathrm{min}}~\underset{\disc}{\mathrm{max}}~\underbrace{\mathbb{E}_{\img} {\left\lVert \img - \gen(\img)  \right\rVert}^{2}_{2}}_{\lossLtwo} + 
       \wadv \cdot \lossgan(\gen,\disc)
\end{equation}
where the weight of the adversarial loss is set as $\wadv=10^{-3}$ in our tests if not stated otherwise.

\section{Experiments}
\label{sec:experiments}
\begin{figure*}[ht]
  \begin{minipage}[b]{0.32\linewidth}
    \centering
      \centerline{\includegraphics[width=\linewidth]{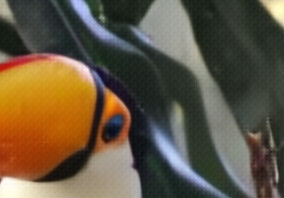}}
       
  \end{minipage}
  \begin{minipage}[b]{0.32\linewidth}
    \centering
      \centerline{\includegraphics[width=\linewidth]{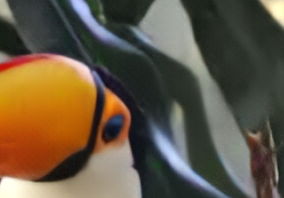}}
       
  \end{minipage}
  \begin{minipage}[b]{0.32\linewidth}
    \centering
      \centerline{\includegraphics[width=\linewidth]{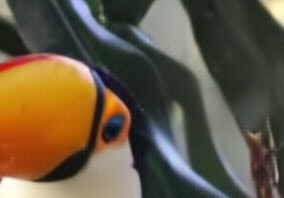}}
       
  \end{minipage}

  \begin{minipage}[b]{0.32\linewidth}
    \centering
    \centerline{\includegraphics[width=\linewidth]{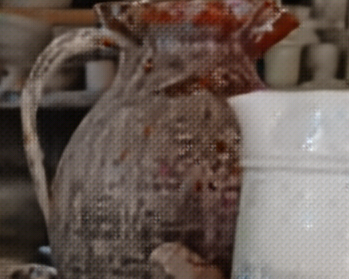}}
       
  \end{minipage}
  \begin{minipage}[b]{0.32\linewidth}
    \centering
    \centerline{\includegraphics[width=\linewidth]{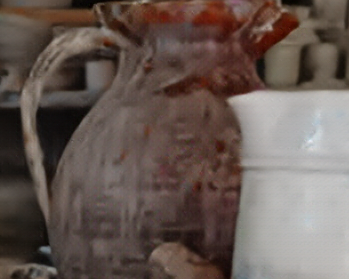}}
       
  \end{minipage}
  \begin{minipage}[b]{0.32\linewidth}
    \centering
    \centerline{\includegraphics[width=\linewidth]{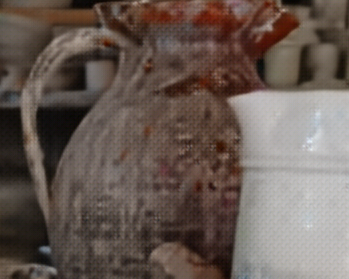}}
       
  \end{minipage}
 
  \begin{minipage}[b]{0.32\linewidth}
    \centering
      \centerline{\includegraphics[width=\linewidth]{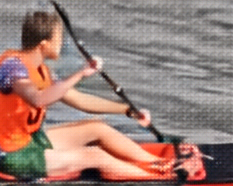}}
       \centerline{Base codec}
  \end{minipage}
  \begin{minipage}[b]{0.32\linewidth}
    \centering
      \centerline{\includegraphics[width=\linewidth]{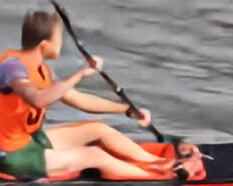}}
       \centerline{Finetuned codec}
  \end{minipage}
  \begin{minipage}[b]{0.32\linewidth}
    \centering
      \centerline{\includegraphics[width=\linewidth]{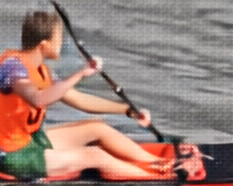}}
       \centerline{Limitedly finetuned codec (\textit{LI})}
  \end{minipage}
\caption[]{The codec finetuned with PatchGAN \textit{(middle)} effectively removes the checkerboard artifacts commonly found in the decoded images of the NN-based convolutional codec such as the base model \textit{(left)}, while the codec finetuned with limited adversarial impact \textit{(right)} only mildly suppresses the artifacts. More examples available at: \url{https://flysofast.github.io/human-finetuned-icm/.}}
\label{fig:zoomed-images}
\end{figure*}

\subsection{Experimental setup}
\label{ssec:exp-setup}
\newparagraph{Dataset and task evaluation:} We used the Open Images V6 \cite{openimages} dataset of 9.2M high quality JPEG images for training and evaluation. The training and finetuning processes of the base model saved the model weights in subsequent checkpoints. Every checkpoint was trained with random 6000 images from the training split of the dataset. Each image was randomly cropped to a $512 \times 512$ patch. The task performance measured in mAP\footnote{\url{https://storage.googleapis.com/openimages/web/evaluation.html}} was obtained by following the Instance Segmentation testing conditions in \cite{vcm_framework}, which consists of tailored benchmarks for the practical use cases of MPEG Video Coding for Machines (VCM). In addition, we also carried out evaluations on other common visual metrics, namely PSNR, SSIM \cite{ssim_wang2004image} and VGG19 \cite{vgg} perceptual loss\footnote{\label{tvmodel}\url{https://pytorch.org/docs/stable/torchvision/models.html}}\footnote{Same as proxy loss in \cite{icme_paper} but apply on the 2nd and 5th layers.}.
\newparagraph{Training:} \textit{The base codec} was trained following the strategy proposed in \cite{icassp_paper} for 200 checkpoints. The proxy loss $\lossproxy$ used the FPN backbone of the MaskRCNN \cite{maskRCNN} network\textsuperscript{\ref{tvmodel}} to extract the feature maps at the top-down pathway. \textit{The PatchGAN-based finetuning} experiments were performed on different settings of patch sizes ($64 \times 64$, $128 \times 128$), number of patches (1, 3, 5) per image and learning rates ($2 \times 10^{-5}$, $2 \times 10^{-9}$). We empirically found that finetuning only the first 2 layers of the decoder sufficiently removed the artifacts effect whilst keeping the number of updated parameters to minimum, which led to reducing the time and resources for the finetuning. We finetuned the codec with Adam \cite{adam} optimizer for 50 checkpoints in each experiments, evaluated the task performance every 10 checkpoints, and evaluated all other metrics on the checkpoint with the highest task score in \autoref{tab:main_results}. Using a batch size of 4, the finetuning process took $\approx$ 7 hours per settings on an NVIDIA A100 GPU.
\begin{table}[ht]
  \caption{Evaluation results of PatchGAN finetuning configurations on different metrics. $\uparrow$ denotes the higher the better and $\downarrow$ denotes the lower the better. The average bitrate of outputs is 0.057 bits per pixel. Configuration with low adversarial impact settings are marked with ``\textit{LI}''}
  \label{tab:main_results}
  \resizebox{\linewidth}{!}{
  \begin{tabular}{@{}p{0.39\linewidth}llll@{}}
  \toprule
\textbf{Configuration}                   & \textbf{mAP}$\uparrow$   & \textbf{PSNR}$\uparrow$   & \textbf{SSIM}$\uparrow$  & \textbf{VGG}$\downarrow$ 
  \\ \midrule
  Base codec                            & \textbf{0.766}          & 27.695          & 0.709          & 0.495           
  \\ \midrule
  3 patches (64x64)                   & 0.752          & 28.526          & 0.764          & 0.497           
  \\
  3 patches (32x32)                   & 0.750          & 28.316          & 0.740          & 0.537           
  \\
  5 patches (64x64)                   & 0.749          & \textbf{28.540}          & \textbf{0.765}          & 0.499           
  \\
  1 patch (64x64)                     & 0.748          & 28.518          & 0.763          & 0.502           
  \\ 
  3 patches (128x128)                 & 0.754          & 28.498          & 0.763          & 0.502           
  \\
  3 patches (128x128), \textit{LI}          & \textbf{0.766} & 28.129          & 0.726          & \textbf{0.469}  
  \\ \bottomrule
  \end{tabular}
 }
\end{table}

\subsection{Experimental results}
\label{ssec:exp-results}
\autoref{tab:main_results} summarises the best task performing checkpoints of the configurations. Overall, all of the configurations can considerably increase PSNR and SSIM scores, which are the metrics that are closely correlated to human perception. The minor differences between the lower rows indicate the robustness of the method. Patch size of $32 \times 32$ shows less appealing results than other sizes in all metrics, possibly due to limited context exposure that hinders the finetuning process. Marginal task performance losses are observed when applying the proposed finetuning in most cases. Therefore, we carried out experiments where the parameter updates due to the adversarial training were greatly attenuated, specifically by reducing $\wadv$ to $1 \times 10^{-4}$ and the learning rate to $2 \times 10^{-9}$. The result shown on the last row of \autoref{tab:main_results} proves that the visual metrics can be improved by the proposed technique without losing task performance. 

\autoref{fig:zoomed-images} shows the close-ups of a few output examples of the compared codecs. We chose the outputs of 2 configurations that resulted in the closest task performance scores to the base codec for the visual comparison. The outputs of the base codec show the typical repetitive patterns found in decoded images of ICM codecs. After being finetuned, the artifacts are almost completely removed, while the image content is well-preserved without any clear random hallucinations. 
The codec that was finetuned with an attenuated adversarial-training could not completely remove the checkerboard patterns, however managed to prevent the task performance loss while improving the scores of visual metrics.

\section{Conclusions}
\label{sec:summary}
In this paper, we propose a novel codec finetuning scheme that can effectively eliminate the well-known ``checkerboard'' pattern produced by deep CNN autoencoders for ICM codecs. Our experimental results demonstrate that the artifacts can be sufficiently removed from the outputs of an ICM codec, thus greatly enhance the visual quality and reconstruction fidelity without any additional components or bitcost. 
Although the machine task accuracy may be affected by this process, the relative performance loss of $-1.6\%$ is negligible in considered cases. 
Furthermore, our study shows that the task performance loss could be totally avoided by confining the impact of the adversarial dynamics. In this settings, all of the benchmarked metrics are improved, suggesting more desirable results could be achieved with granularity in training setup. We also found that only a small subset of the codec parameters is responsible for the artifacts, future study on this subject could lead to interesting findings in neural networks explainability.

\bibliographystyle{template/IEEEbib}
\bibliography{refs}

\end{document}